# Biological, Family and Cultural Predictors of Personality Structure analysis based on personality prediction models constructed by open data source

Cheng Hua[1] ,Wang Dandan[2]

# Abstract

**Objective:** This study takes further step on understanding personality structure in order to cope with the mental health during the COVID-19 global pandemic situation.

**Methods:** Categorized the independent variables into biological, family and cultural predictors according to the datasets of the Big-5 personality survey online. And established multiple regression prediction models and exhaustive CHAID decision tree model of each personality trait.

**Results:** Females are different from males in personality. The personality changes when growing. One-handed dominants are less agreeable and open than those who use both hands. Different sexual orientation does have variety personality. Native language used and education attainment is significantly related to personality accordingly. Marriage did help shaping personality to be more extroverted, less neurotic or agreeable and more conscientious and open. People raised in urban are more agreeable and open. Neurotic and open people often come from small families. person participated in voting are more extroverted, conscientious and open but less neurotic and agreeable. Different religions and races have different characteristics in each dimension of personality and there is no clear pattern have been found.

**Conclusion:** Personality traits are indeed affected by multiple confounding factors. but the exploration on multiple cultures predictors still needed more details.

# Keywords

Big Five Personality traits; Multivariate Linear Regression Model; Exhaustive CHAID Decision Tree Model; Gender Differences; Cross-age Differences; Hand Preference; Sexual Orientation; Family and culture factors.

[1] : Corresponding Author : Cheng Hua M.D., Llingnan Normal University, Sport Science Institute, Zhanjiang, Guangdong, China. Email: trernghwhuare@aliyun.com. Tel: +86-13828217170. ORCID iD: https://orcid.org/0000-0003-0903-6766.

[2] : author: Wang Dandan, Guangdong Medical University, Department of Psychology, Zhanjiang, Guangdong, China.

# Introduction

Understanding personality and abnormal behavior has been intriguing to scholars since before 3000 BC. Scientists learned human personality by capturing and organizing the behavior and emotion differences individually during the era where technological development is far from being developed today. Five independent dimensions of personality variation, known as the Big Five factors, originally discovered in investigations of English-language personality-descriptive adjectives. Researchers already reached a consensus that five broad and roughly independent dimensions are the best summary of personality domains. The names of these Big Five or Five-Factor Model (B5/FFM) are extraversion(E), neuroticism(N), openness(O), agreeableness(A) and conscientiousness(C), are believed to capture the normative personality[1].

A quantitative review stated that for the diagnostic specific depressive, anxiety and substance use disorders were high on N and low on C. while A and O were largely unrelated to the analyzed diagnoses. And many dysthymic disorder and social phobia showed low E[2]. Nowadays the COVID-19 pandemic could be a significant risk factor for people's mental diseases around the world[3].And personality traits (pts) provide an exploration in intervention, treatments and prevention activities for post-traumatic symptom patterns.

Pts are often emphasized to be as equally important for many aspects of life as cognitive ability for predicting social and economic success base on a large body of evidence. Gender differences were found in adolescent along the timing of personality maturation[4]. And in western, educated, industrialized, rich, and democratic populations, pts of N with O but not C is a strong predictor income, which is different from the previous literature has found C and N to be the strongest predictors in low- and middle-income countries [5]. Across-culture consistency of big-5 structure challenge in indigenous traits in other languages, because the extent to endorsed pts varies considerably in each culture[6].

In this study, we perform statistical analysis of available large data sets from web-based questionnaires of global populations with the big-5 structure in personality traits through three samples with cross-cultural, ethnic, religious, gender, language and various bio-social factors, in order to provide references for mental health protection strategies during the COVID-19 pandemic.

# Methods

All raw data collected anonymously through Open-Source Psychometrics Project complying with the privacy protection policy (https://openpsychometrics.org/privacy_policy/). Users were informed well that their answers were stored and analyzed for scientific purposes and asked to confirm their answers were accurate and suitable for research. There are 3 major datasets were collected in this study, named as Sample 1~3 ( Sample 1: http://openpsychometrics.org/_rawdata/BIG5.zip; Sample 2: https://openpsychometrics.org/_rawdata/duckworth-grit-scale-data.zip; and Sample 3: https://openpsychometrics.org/_rawdata/IPIP-FFM-data-8Nov2018.zip ).

This study was confirmed by the Ethics Committee of Institutional Ethics Committee of Lingnan

Normal university (Approval No: 001).

## Measuring tool

Personality traits measuring by the big-5 personality traits using the self-report International Personality Item Pool (IPIP) Big-5 Factor Markers. The big-5 personality traits developed by Goldberg[7] are the most accepted and commonly used personality models in academic psychology. The Big Five personality stands out in explaining the answers to personality questions: extraversion, neuroticism, agreeableness, conscientiousness and openness.

## Data collecting and processing

Available data sets include thousands of Answers to the Big Five Personality Test of web-based questionnaires constructed with items from the IPIP.
First calculate the scores of every trait of the big-5 personality for each participant according to the formula below. The sum of each items' scores, including +keyed and -keyed items, equals the specific feature's factor score.

$$Y_m = \sum_{9}^{1}[X_n + (6 - X_{n+1})] \tag{1}$$

In this formula, $Y$ is the total score of the trait factor; $m$ stands for 5 traits factors, which are E, N, A, C and O. $X_n$ is the score of each item. $n= 1\sim9$. The total number of factor entries in this category is 10.
Secondly, the independent variables are classified and assigned values in hierarchical order, and all missing variables are assigned as zero. Then extract and classify to biological, family and culture factor variables in the data set(see *Tab*.1).

Tab. 1 extracted independent variables and their ordered classification

| VARIABLE | ORDINAL CATEGORY |
|---|---|
| *Growth* | (1) youth=12~24;(2) early adulthood=25~40; (3) late adulthood=41~60; (4) old age=>61 |
| *Gender* | (1) Male; (2) Female; (3) Other. |
| *Hand* [a] | (1) Right, (2) Left; (3) Both. |
| *Education* | (1) Less than high school; (2) High school; (3) University degree; (4) Graduate degree. |
| *Urban* [b] | (1) Rural (country side); (2) Suburban; (3) Urban (town, city). |
| *Gender* | (1) Male; (2) Female; (3) Other. |
| *Engnat* [c] | (1) Yes; (2) No. |
| *Hand* [d] | (1) Right, (2) Left; (3) Both. |
| *Voted* [e] | (1) Yes; (2) No. |
| *Married* | (1) Never married; (2) Currently married; (3) Previously married. |
| *growth* | (1) youth=12~24;(2) early adulthood=25~40; (3) late adulthood=41~60; (4) old age=>61 |
| *Religion* | (1) Agnostic; (2) Atheist; (3) Buddhist; (4) Christian (Catholic); (5) Christian (Mormon); (6) Christian (Protestant); (7) Christian (Other); (8) Hindu; (9) Jewish; (10) Muslim; (11) Sikh; (12) Other. |
| *Orientation* | (1) Heterosexual; (2) Bisexual; (3) Homosexual;(4) Asexual; (5) Other. |
| *Race* | (1) Asian; (2) Arab; (3) Black; (4) Indigenous Australian Native, American or White; (5) Other |
| *Family* | (1) small (1~3) ;(2) medium (4~10) ;(3) large (>=11) |

[a]: "hand" means dominant hand. i.e., Response to "What hand do you use to write with?"

[b]: "urban" means environment of being raised. i.e., Response to "What type of area did you live when you were a child?"

[c]: "engnat" is short for English native speaker. i.e., Response to "is English your native language?"

[d]: "hand" means dominant hand. i.e., Response to "What hand do you use to write with?"

[e]: "voted" is the answer to "Have you voted in a national election in the past year?"

The third step is to integrate data with the same variables from data sets 1 and 2 to form a new data set in order to reduce the error between data sets. The dependent variables are continuous variable and the independent variables are categorical variables in the new data set (see *Tab*.2).

Tab. 2 assignments and value ranges of valuables

| | continuous variables | Assignment and value range |
|---|---|---|
| *continuous variables* ($Y_m$) | Extroversion | $Y_E \in [10,50]$ |
| | Neuroticism | $Y_N \in [10,50]$ |
| | Agreeableness | $Y_A \in [10,50]$ |
| | Conscientiousness | $Y_C \in [10,50]$ |
| | Openness | $Y_O \in [10,50]$ |
| *categorical variables* ($X_i$) | Gender | $X_{Gender} = (1,2,3)$ |
| | Hand | $X_{Hand} = (1,2,3)$ |
| | growth | $X_{Growth} = (1,2,3,4)$ |
| | Education | $X_{Education} = (1,2,3,4)$ |
| | Urban | $X_{Urban} = (1,2,3)$ |
| | Orientation | $X_{Orientation} = (1,2,3,4,5)$ |
| | Engnat | $X_{Engnat} = (1,2)$ |
| | Married | $X_{Married} = (1,2,3)$ |
| | Family | $X_{Family} = (1,2,3)$ |
| | Voted | $X_{Voted} = (1,2)$ |
| | Religion | $X_{Religion} = (1,2,\ldots,12)$ |
| | Race | $X_{Race} = (1,2,3,4,5)$ |

# Statistic

To answer the question how to predict $Y_m$ on the basis of $X_i$ or to describe how continuous variables $Y_m$ depends on the categorical variables $X_i$, multiple linear regression analysis was applied to accomplished the questions of interest in this study.

$$Y_m = \beta_0 + \beta_1 X_1 + \cdots + \beta_\rho X_\rho \tag{2}$$

In *Equ*.2, $\beta_0$ is intercept. $\beta_1 \cdots \beta_\rho$ are the regression coefficients. $X_i$ $(X_1, \cdots, X_\rho)$ is defined as the independent variable as well as the predictor of the regression model. And $Y_m(Y_E, Y_N, Y_A, Y_C, Y_O)$ is the dependent variable which is also the outcome of the model. In this study, the independent variables are divided into biological parameters ($X_{Gender}$, $X_{Growth}$, $X_{Hand}$), family factors ($X_{Education}$, $X_{Urban}$, $X_{Engnat}$, $X_{Orientation}$, $X_{Married}$, $X_{Family}$) and culture factors ($X_{religion}$, $X_{race}$, $X_{voted}$).

Factors' prediction models were constructed also by applying Exhaustive CHAID. Exhaustive CHAID is an improved algorithm of *Chi-squared Automatic Interaction Detector* method based on the $\chi^2$ test as the criterion. The decision tree growth process is divided into merging steps and splitting steps in method. The merging process is to first perform a $\chi^2$ test on the correlation between each independent variable and the dependent variable, examine the differences between all levels, and finally form two large levels, and then group the two large levels to form a difference level group. The splitting process examines the correlation between the reclassified independent variables and the dependent variables, and determines the order in which the independent variables enter the decision tree. The correlations are entered from high to low until there are significant differences and no need to continue splitting.

## Results

The total participants' count of three samples are Sample 1 (N=19708), sample 2 (N=3874), and sample 3 (N=874434). $\bar{M} \pm \mathbf{SD}$ of 3 datasets in sample 1 ($\bar{Y_E}$=30.11±9.22, $\bar{Y_N}$=31.74±3.79, $\bar{Y_A}$=23.44±5.79, $\bar{Y_C}$=32.20±6.48, $\bar{Y_O}$=36.66±6.98), sample 2 ($\bar{Y_E}$=30.19±9.15, $\bar{Y_N}$=32.06±3.94, $\bar{Y_A}$=23.41±5.92, $\bar{Y_C}$=32.30±6.54, $\bar{Y_O}$=36.44±4.01) and sample 3 ($\bar{Y_E}$=29.59±9.10, $\bar{Y_N}$=31.96±3.83, $\bar{Y_A}$=23.88±5.87, $\bar{Y_C}$=32.41±6.59, $\bar{Y_O}$=36.58±4.00). The differences of each trait factor of the three samples are not statistically significant (*F*=0.00, *P*=1.00). Namely, the factor structures from 3 different data sets are quite similar.

In the comparison of $\bar{M} \pm \mathbf{SD}$, it can be concluded that the as the independent variables vary, the dependent variables change accordingly, especially in $X_{Gender}$, $X_{Growth}$, $X_{Education}$, $X_{Orientation}$, $X_{Married}$, $X_{Family}$, $X_{Religion}$, $X_{Race}$ and $X_{Voted}$ (all $P<0.05$), as *Tab*.3 presented. Take the sample 3 for example, it is a dataset of composed of 874434 valid participants from 225 countries. The differences in the $\bar{M} \pm \mathbf{SD}$ of every big-5-structure of personality trait are statistically significant (all *P*=0.00) from one country to another (*Fig*.1).

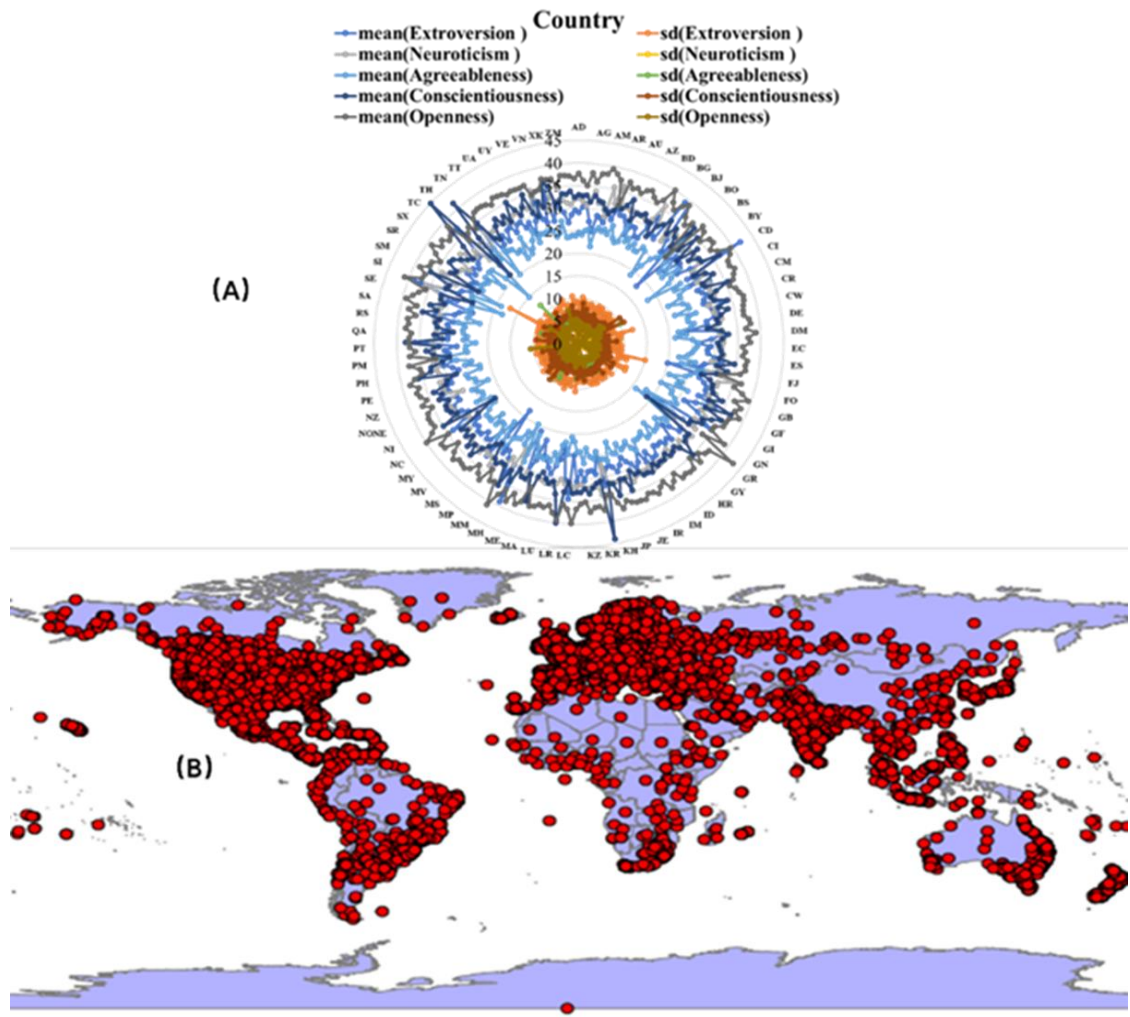

Fig. 1 Descriptive Statistics of participants from sample 3

(A) Radar chart and of the subscale score $\bar{M} \pm \mathbf{SD}$ of the big-5-personlity questionnaire from 225 countries; (B) map visualization of Global participants.

For example, compare cases between Canada (country=CA) and the United States (country=US). Although there are large differences in the calculation of participants, the $\bar{M} \pm \mathbf{SD}$ scores of each trait factor are relatively different and statistically significant ($P$=0.00). The profile of the factor distribution graph is similar (*Fig.2*).

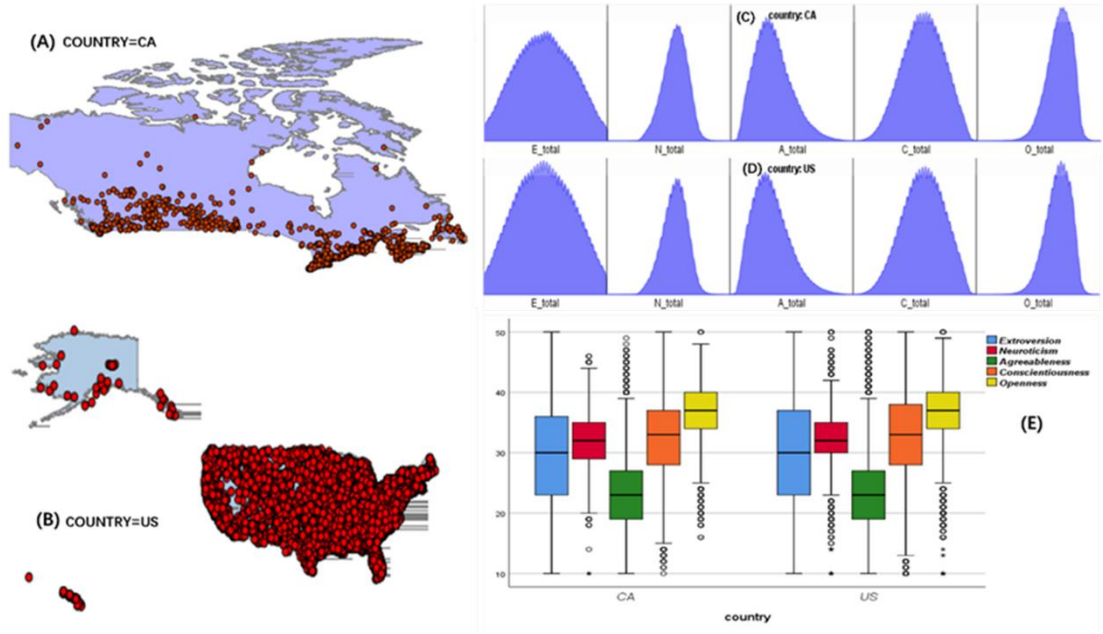

Fig. 2 Descriptive Statistics of Participants of the big-5-personality inventory online from US and CA (A) map visualization of participant from CA;(B) map visualization of participant from US;(C) Distribution charts of the $\bar{M} \pm SD$ of each personality trait score in CA;(D) Distribution charts of the $\bar{M} \pm SD$ of each personality trait score in US;(E) Box plot of Comparison of 5 personality traits scores between CA and US participants.

## Multivariate linear regression models for personality trait prediction

Apply regression equation model to test how the biological factors ($X_{Growth}, X_{Gender}, X_{Hand}$), the family factors ($X_{Education}$, $X_{Urban}$, $X_{Engnat}$, $X_{Orientation}$, $X_{Married}$, $X_{Family}$) and the culture factors ($X_{religion}, X_{race}, X_{voted}$) predict personality traits ($Y_m$). The three sets of equations are as follows (*Equ.* 3~7, *Equ.* 8~12 and *Equ.* 13~17).

***Predicted by biological factors***

$$Y_E = 27.758 + 1.153 X_{Growth} + 0.637 X_{Gender} - 0.378 X_{Hand} \qquad (3)$$
$$Y_N = 30.405 - 0.434 X_{Growth} + 1.294 X_{Gender} - 0.045 X_{Hand} \qquad (4)$$
$$Y_A = 28.585 - 1.268 X_{Growth} - 2.278 X_{Gender} + 0.444 X_{Hand} \qquad (5)$$
$$Y_C = 29.105 + 1.607 X_{Growth} + 0.410 X_{Gender} - 0.019 X_{Hand} \qquad (6)$$
$$Y_O = 36.143 + 0.538 X_{Growth} - 0.434 X_{Gender} + 0.310 X_{Hand} \qquad (7)$$

In *Equ.*3, the regression coefficient for $X_{Growth}$(1.153) means that the estimated mean increase in $Y_E$ increase 1.153 if $X_{Growth}$ increase one unit, while $X_{Gender}$ and $X_{Hand}$ remain constant. This means that as aging, the score for extroversion also increases in phases. The coefficient for $X_{Gender}$(0.637) means that if $X_{Gender}$ increases by 1 unit, $Y_E$ Increases by 0.637. That is, females score 0.637 higher than males. The estimated coefficient for $X_{Hand}$(-0.378) means if $Y_E$ Decrease 0.378 if $X_{Hand}$ increase one unit.

This set of regression equations suggests that as aging, $Y_E$, $Y_C$, $Y_O$ score increase and $Y_N$, $Y_A$ decrease. Under the same conditions, females scored higher in $Y_E$, $Y_N$, and $Y_C$ than males, while males scored higher in $Y_A$ and $Y_O$ than females. People who are accustomed to one-handed score higher on the $Y_E$, $Y_N$ and $Y_C$; those who are accustomed to both hands score higher on the $Y_A$ and $Y_O$.

The relationship between other personality trait factors and biological parameters can be derived in the same way. All factor regression models are statistically significant (P=0.00) (See *Tab*.3).

Tab. 3 multivariate linear regression models $Y_m$ predicted by biological factors ($X_{Growth}$, $X_{Gender}$, $X_{Hand}$)

| Model | | B | SE | T | P | F | P | Equation |
|---|---|---|---|---|---|---|---|---|
| $Y_E$ | Constant | 27.758 | 0.295 | 94.084 | 0.000 | 78.783 | 0.000 | (3) |
| | $X_{Gender}$ | 1.153 | 0.080 | 14.379 | 0.000 | | | |
| | $X_{Hand}$ | 0.637 | 0.121 | 5.273 | 0.000 | | | |
| | $X_{Growth}$ | -0.378 | 0.145 | -2.606 | 0.009 | | | |
| $Y_N$ | Constant | 30.405 | 0.121 | 252.244 | 0.000 | 298.321 | 0.000 | (4) |
| | $X_{Gender}$ | -0.434 | 0.033 | -13.245 | 0.000 | | | |
| | $X_{Hand}$ | 1.294 | 0.049 | 26.239 | 0.000 | | | |
| | $X_{Growth}$ | -0.045 | 0.059 | -0.765 | 0.445 | | | |
| $Y_A$ | Constant | 28.585 | 0.181 | 157.660 | 0.000 | 524.030 | 0.000 | (5) |
| | $X_{Gender}$ | -1.268 | 0.049 | -25.728 | 0.000 | | | |
| | $X_{Hand}$ | -2.278 | 0.074 | -30.709 | 0.000 | | | |
| | $X_{Growth}$ | 0.444 | 0.089 | 4.983 | 0.000 | | | |
| $Y_C$ | Constant | 29.105 | 0.205 | 141.696 | 0.000 | 280.849 | 0.000 | (6) |
| | $X_{Gender}$ | 1.607 | 0.056 | 28.787 | 0.000 | | | |
| | $X_{Hand}$ | 0.410 | 0.084 | 4.877 | 0.000 | | | |
| | $X_{Growth}$ | -0.019 | 0.101 | -0.193 | 0.847 | | | |
| $Y_O$ | Constant | 36.143 | 0.128 | 283.257 | 0.000 | 115.131 | 0.000 | (7) |
| | $X_{Gender}$ | 0.538 | 0.035 | 15.529 | 0.000 | | | |
| | $X_{Hand}$ | -0.434 | 0.052 | -8.312 | 0.000 | | | |
| | $X_{Growth}$ | 0.310 | 0.063 | 4.945 | 0.000 | | | |

***Predicted by family factors***

$$Y_E = 29.048 + 0.708 X_{Education} + 0.307 X_{Urban} - 0.944 X_{Engnat} - 0.790 X_{Orientation} + 0.758 X_{Married} - 0.172 X_{Family} \quad (8)$$
$$Y_N = 34.724 - 0.343 X_{Education} - 0.040 X_{Urban} - 0.586 X_{Engnat} + 0.081 X_{Orientation} - 0.483 X_{Married} - 0.387 X_{Family} \quad (9)$$
$$Y_A = 23.811 - 0.722 X_{Education} + 0.035 X_{Urban} + 1.597 X_{Engnat} + 0.320 X_{Orientation} - 0.774 X_{Married} - 0.301 X_{Family} \quad (10)$$
$$Y_C = 27.866 + 0.942 X_{Education} + 0.138 X_{Urban} - 0.545 X_{Engnat} - 0.306 X_{Orientation} + 1.519 X_{Married} + 0.927 X_{Family} \quad (11)$$
$$Y_O = 35.405 + 0.634 X_{Education} - 0.083 X_{Urban} - 0.246 X_{Engnat} + 0.049 X_{Orientation} + 0.199 X_{Married} - 0.231 X_{Family} \quad (12)$$

In *Equ*.8, the estimated mean increase in $Y_E$ increase 0.71 if $X_{Education}$ increase one unit, while other predictors remain constant. It can be inferred that the predictive effect of the regression coefficient of each other predictor variable on the regression model.

This set of regression equations suggest that as the level of education increases, the scores of personality dimensions in the $Y_E, Y_C$, and $Y_O$ will increase, and the scores in the $Y_N$ and $Y_A$ will decrease. it can be Explained that education shapes the personality to become more extroverted, more conscientious and more open to experience, but less neurotic and agreeable to some extent. Children who grew up in urban might probably get higher scores in $Y_E, Y_A$, and $Y_C$, and lower $Y_N$ and $Y_O$. They are more extorverted, agreeable and conscientious but less neurotic and open. Native English speakers have higher scores on the $Y_E, Y_N, Y_C$, and $Y_O$ and lower scores on the $Y_A$ than non-native English speakers. That is, native English speakers are more extraverted, neurotic, conscientious and open, but their personality is not as agreeable as non-native English speakers. Heterosexuals score higher in the $Y_E$ and $Y_C$ dimensions, and asexuals score higher in the $Y_N, Y_A$, and $Y_O$. Marriage help person more extraverted, conscientious and open, less neurotic, agreeable. Person comes from small family are more extraverted, neurotic, agreeable and open but less conscientious.

All family factor regression models are statistically significant (*P*=0.00) (See *Tab*.4).

Tab. 3  multivariate linear regression models $Y_m$ predicted by family factors ($X_{Education}$, $X_{Urban}$, $X_{Engnat}$, $X_{Orientation}$, $X_{Married}$, $X_{Family}$)

| MODEL | | B | SE | T | P | F | P | Equation |
|---|---|---|---|---|---|---|---|---|
| $Y_E$ | Constant | 29.408 | 0.821 | 35.810 | 0.000 | | | |
| | $X_{Education}$ | 0.708 | 0.168 | 4.210 | 0.000 | | | |
| | $X_{Urban}$ | 0.307 | 0.188 | 1.631 | 0.103 | | | |
| | $X_{Engnat}$ | -0.944 | 0.314 | -3.012 | 0.003 | 12.354 | .000 | (8) |
| | $X_{Orientation}$ | -0.790 | 0.149 | -5.307 | 0.000 | | | |
| | $X_{Married}$ | 0.758 | 0.284 | 2.674 | 0.008 | | | |
| | $X_{Family}$ | -0.172 | 0.323 | -0.531 | 0.596 | | | |
| $Y_N$ | Constant | 34.724 | 0.353 | 98.436 | 0.000 | | | |
| | $X_{Education}$ | -0.343 | 0.072 | -4.743 | 0.000 | | | |
| | $X_{Urban}$ | -0.040 | 0.081 | -0.498 | 0.619 | | | |
| | $X_{Engnat}$ | -0.586 | 0.135 | -4.349 | 0.000 | 15.676 | .000 | (9) |
| | $X_{Orientation}$ | 0.081 | 0.064 | 1.258 | 0.208 | | | |
| | $X_{Married}$ | -0.483 | 0.122 | -3.962 | 0.000 | | | |
| | $X_{Family}$ | -0.387 | 0.139 | -2.785 | 0.005 | | | |
| $Y_A$ | Constant | 23.811 | 0.525 | 45.348 | 0.000 | | | |
| | $X_{Education}$ | -0.722 | 0.108 | -6.713 | 0.000 | | | |
| | $X_{Urban}$ | 0.035 | 0.120 | 0.290 | 0.772 | | | |
| | $X_{Engnat}$ | 1.597 | 0.201 | 7.962 | 0.000 | 27.702 | .000 | (10) |
| | $X_{Orientation}$ | 0.320 | 0.095 | 3.363 | 0.001 | | | |
| | $X_{Married}$ | -0.774 | 0.181 | -4.267 | 0.000 | | | |
| | $X_{Family}$ | -0.301 | 0.207 | -1.457 | 0.145 | | | |
| $Y_C$ | Constant | 27.866 | 0.577 | 48.318 | 0.000 | | | |
| | $X_{Education}$ | 0.942 | 0.118 | 7.969 | 0.000 | | | |
| | $X_{Urban}$ | 0.138 | 0.132 | 1.043 | 0.297 | | | |
| | $X_{Engnat}$ | -0.545 | 0.220 | -2.474 | 0.013 | 36.042 | .000 | (11) |
| | $X_{Orientation}$ | -0.306 | 0.105 | -2.929 | 0.003 | | | |
| | $X_{Married}$ | 1.519 | 0.199 | 7.626 | 0.000 | | | |
| | $X_{Family}$ | 0.927 | 0.227 | 4.085 | 0.000 | | | |
| $Y_O$ | Constant | 35.405 | 0.359 | 98.639 | 0.000 | | | |
| | $X_{Education}$ | 0.634 | 0.074 | 8.616 | 0.000 | | | |
| | $X_{Urban}$ | -0.083 | 0.082 | -1.015 | 0.310 | | | |
| | $X_{Engnat}$ | -0.246 | 0.137 | -1.793 | 0.073 | 16.255 | .000 | (12) |
| | $X_{Orientation}$ | 0.049 | 0.065 | 0.758 | 0.448 | | | |
| | $X_{Married}$ | 0.199 | 0.124 | 1.608 | 0.108 | | | |

|   | $X_{Family}$ | -0.231 | 0.141 | -1.638 | 0.101 |   |

*Predicted by culture factors*

$$Y_E = 28.946 + 0.214X_{religion} + 0.423X_{race} - 0.768X_{voted} \quad (13)$$
$$Y_N = 30.807 - 0.063X_{religion} + 0.154X_{race} + 0.623X_{voted} \quad (14)$$
$$Y_A = 24.198 - 0.178X_{religion} - 0.349X_{race} + 0.785X_{voted} \quad (15)$$
$$Y_C = 33.859 + 0.150X_{religion} + 0.049X_{race} - 1.480X_{voted} \quad (16)$$
$$Y_O = 36.784 - 0.042X_{religion} + 0.252X_{race} - 0.589X_{voted} \quad (17)$$

Different religions and racial cultures population groups various in personality dimensions. People who voted are more extroverted, conscientious, and open. And person didn't vote are more neurotic and agreeable. and vice versa(see *Tab.*5).

Tab. 4 multivariate linear regression models $Y_m$ predicted by culture factors ($X_{religion}$, $X_{race}$, $X_{voted}$)

| MODEL |   | B | SE | T | P | F | P | Equation |
|---|---|---|---|---|---|---|---|---|
| $Y_E$ | Constant | 28.946 | 0.710 | 40.790 | 0.000 | 13.201 | 0.000 | (13) |
|   | $X_{religion}$ | 0.214 | 0.044 | 4.860 | 0.000 |   |   |   |
|   | $X_{race}$ | 0.423 | 0.113 | 3.751 | 0.000 |   |   |   |
|   | $X_{voted}$ | -0.768 | 0.304 | -2.527 | 0.012 |   |   |   |
| $Y_N$ | Constant | 30.807 | 0.305 | 100.921 | 0.000 | 15.851 | 0.000 | (14) |
|   | $X_{religion}$ | -0.063 | 0.019 | -3.325 | 0.001 |   |   |   |
|   | $X_{race}$ | 0.154 | 0.048 | 3.174 | 0.002 |   |   |   |
|   | $X_{voted}$ | 0.623 | 0.131 | 4.764 | 0.000 |   |   |   |
| $Y_A$ | Constant | 24.198 | 0.457 | 52.928 | 0.000 | 23.430 | 0.000 | (15) |
|   | $X_{religion}$ | -0.178 | 0.028 | -6.294 | 0.000 |   |   |   |
|   | $X_{race}$ | -0.349 | 0.073 | -4.802 | 0.000 |   |   |   |
|   | $X_{voted}$ | 0.785 | 0.196 | 4.008 | 0.000 |   |   |   |
| $Y_C$ | Constant | 33.859 | 0.505 | 67.004 | 0.000 | 23.049 | 0.000 | (16) |
|   | $X_{religion}$ | 0.150 | 0.031 | 4.781 | 0.000 |   |   |   |
|   | $X_{race}$ | 0.049 | 0.080 | 0.616 | 0.538 |   |   |   |
|   | $X_{voted}$ | -1.480 | 0.216 | -6.842 | 0.000 |   |   |   |
| $Y_O$ | Constant | 36.784 | 0.310 | 118.481 | 0.000 | 18.193 | 0.000 | (17) |
|   | $X_{religion}$ | -0.042 | 0.019 | -2.206 | 0.027 |   |   |   |
|   | $X_{race}$ | 0.252 | 0.049 | 5.122 | 0.000 |   |   |   |
|   | $X_{voted}$ | -0.589 | 0.133 | -4.430 | 0.000 |   |   |   |

# Exhaustive CHAID decision tree models for personality trait prediction

Taking continuous variables of $Y_E$, $Y_N$, $Y_A$, $Y_C$ and $Y_O$ as the dependent variables, and the biological categorical variables $X_{Gender}$, $X_{Growth}$ and $X_{Hand}$ as the independent variables, the Exhausted CHAID decision tree personality trait biological parameters prediction models were constructed. Family factors prediction models were constructed by family factors, the categorical variables of $X_{Education}$, $X_{Urban}$, $X_{Engnat}$, $X_{Orientation}$, $X_{Married}$, $X_{Family}$ as the independent variables. And culture factors such as $X_{religion}, X_{race}, X_{voted}$ categorical variables as the independent variables constructed the Exhausted CHAID decision tree personality trait culture factors prediction model. The independent variables included in the results and tree nodes of each model for all factor prediction decision tree models are as follows (see *Tab*.6)

Tab. 5 variables included in the results and tree nodes of each exhaustive CHAID decision tree model

| Models | Biological parameters | | Family factors | | Culture factors | |
|---|---|---|---|---|---|---|
| | factors | Nodes | factors | Nodes | factors | Nodes |
| $Y_E$ | $X_{Gender}, X_{Growth}$ | 0~10 | $X_{Education}, X_{Orientation}, X_{Engnat}, X_{Married}$ | 0~9 | $X_{religion}, X_{race}, X_{voted}$ | 0-8 |
| $Y_N$ | $X_{Gender}, X_{Growth}$ | 0~9 | $X_{Education}, X_{Engnat}, X_{Married}$ | 0~10 | $X_{religion}, X_{voted}$ | 0~5 |
| $Y_A$ | $X_{Gender}, X_{Growth}, X_{Hand}$ | 0~14 | $X_{Education}, X_{Orientation}, X_{Engnat}, X_{Married}$ | 0~9 | $X_{religion}, X_{race}, X_{voted}$ | 0~10 |
| $Y_C$ | $X_{Gender}, X_{Growth}$ | 0~8 | $X_{Education}, X_{Orientation}, X_{Married}$ | 0~12 | $X_{religion}, X_{voted}$ | 0~10 |
| $Y_O$ | $X_{Gender}, X_{Growth}, X_{Hand}$ | 0~11 | $X_{Education}, X_{Orientation}, X_{Engnat}, X_{Family}$ | 0~15 | $X_{religion}, X_{voted}$ | 0~7 |

*Extroversion prediction model*

Females at their late adulthood (Node 9, $\overline{Y_E} = 32.46$) are more extroverted while males of youth (Node 7, $\overline{Y_E} = 29.07$) are not so much extroverted, regardless of which dominant hand (*Fig*.3-A). Previously married person of heterosexual orientation with university or graduate degree (Node 9, $\overline{Y_E} = 34.18$) are more extroverted. Persons of asexual orientation (Node 3, $\overline{Y_E} = 26.19$) are less in extroversion of personality (*Fig*.3-B). Jewish (Node 3, $\overline{Y_E} = 33.83$) regardless voted or not are more extroverted and Atheist who have never voted (Node 5, $\overline{Y_E} = 27.93$) are less extroverted (*Fig*.3-C).

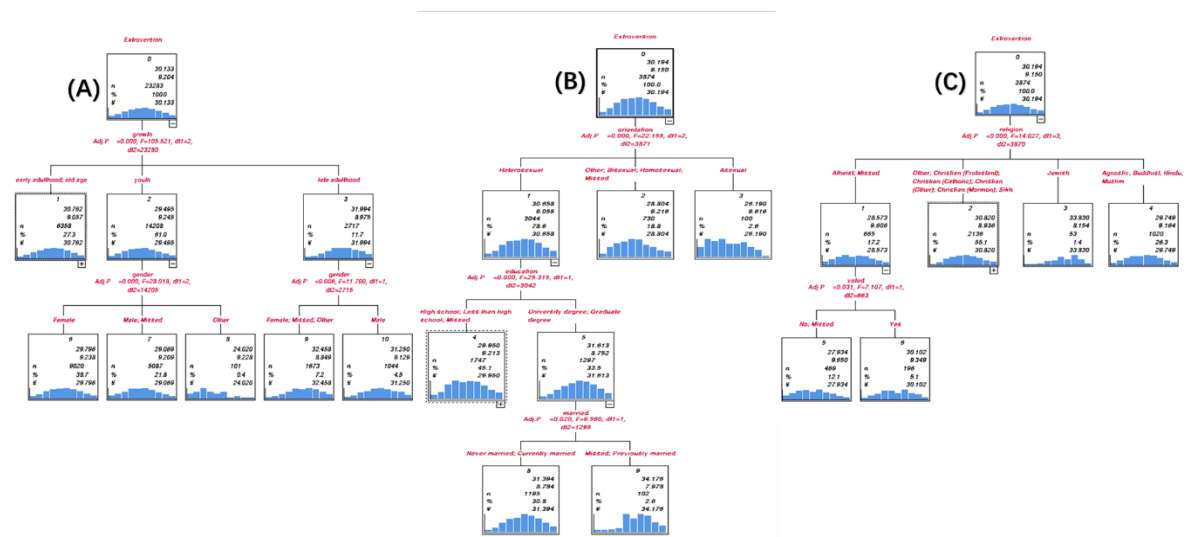

Fig. 3 Exhaustive CHAID decision tree models for E prediction predicted by .
A: biological parameters; B: family factors; C: culture factors.

*Neuroticism prediction model*

Females of youth (Node 4, $\overline{Y_N} = 32.59$) are neurotic while males of late adulthood and old age (Node 7, $\overline{Y_N} = 30.53$) get less score in Neuroticism (*Fig*.4-A). Native English speaker with the education of less than high school (Node 7, $\overline{Y_N} = 32.95$) are more neurotic. And person with graduate degree in education (Node 4, $\overline{Y_N} = 31.16$) are less neurotic regardless of their language. (*Fig*.4-B). Religious people Atheists, Christians (Catholic & Mormon), Jewish and Agnostic not voted (Node 3, $\overline{Y_N} = 32.58$) are more neurotic. people who have voted regardless of their religion or race (Node 2, $\overline{Y_N} = 31.53$) have lower scores on neurotic traits (*Fig*.4-C).

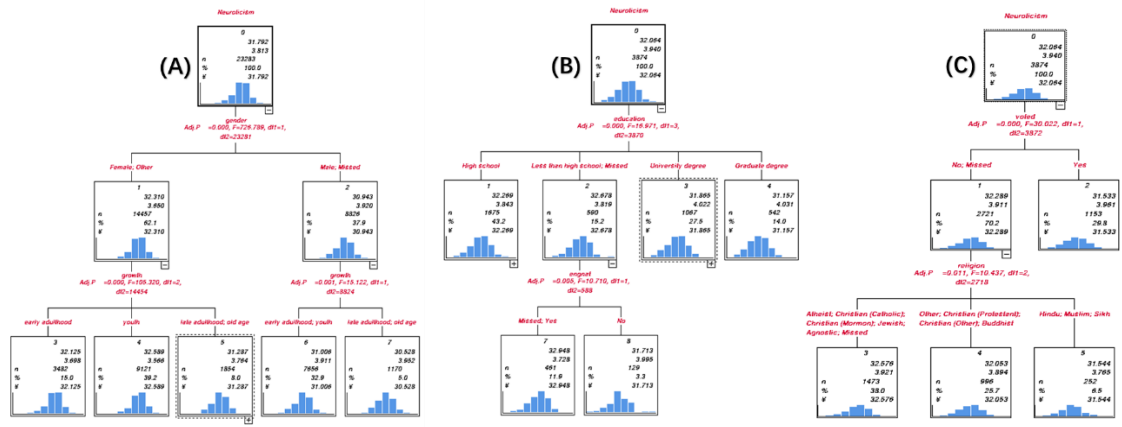

Fig. 4 Exhaustive CHAID decision tree models for N prediction predicted by
A: biological parameters; B: family factors; C: culture factors.

*Agreeableness prediction model*

Both-handed dominants of youth males (Node 13, $\overline{Y_A} = 27.07$) are more agreeable. Females of late adulthood and old age (Node 5, $\overline{Y_A} = 20.42$) are not that agreeable in personality (*Fig*.5-A). Non-native English speaker with high school or less than high school (Node 3, $\overline{Y_A} = 25.39$) in education are more agreeable. And native English speakers with university degree or graduate degree in education who currently married or previously married (Node 12, $\overline{Y_A} = 21.09$) are less agreeable in personality (*Fig*.5-B). The Atheist (Node 1, $\overline{Y_A} = 25.42$) are more agreeable in personality. In the other hand, the religious people of Christians (Protestant, Other, Mormon) from the race of Indigenous, Australian, Native American or White who have voted (Node 10, $\overline{Y_A} = 21.18$) are less agreeable (*Fig*.5-C).

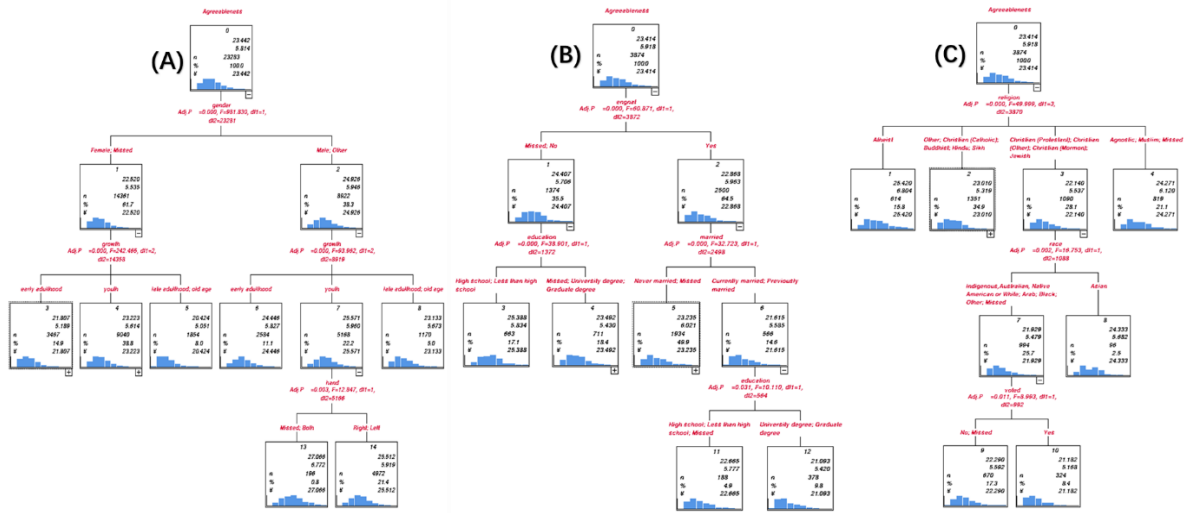

Fig. 5  Exhaustive CHAID decision tree models for A prediction predicted by
A: biological parameters; B: family factors; C: culture factors.

*Conscientiousness prediction model*

People are more conscientious at their late adulthood and old age (Node 3, $\bar{Y}_C = 34.62$) regardless the gender while less conscientious at youth, especially males of youth (Node 7, $\bar{Y}_C = 31.05$) (*Fig*.6-A). Person of heterosexual or asexual orientation who currently married or previously married (Node 5, $\bar{Y}_C = 35.07$) are more conscientious. Bisexual or asexual orientation that never married and with only high school or less than high school in education (Node 11, $\bar{Y}_C = 29.31$) are less in conscientiousness of personality (*Fig*.6-B). Religious people of Christians (Protestant, Mormon & Other) and Sikh who have voted (Node 10, $\bar{Y}_C = 34.83$) are more conscientious in personality. Atheists and agnostics who have not participated in the vote (Node 5, $\bar{Y}_C = 30.44$) do not score high in the conscientious dimension (*Fig*.6-C).

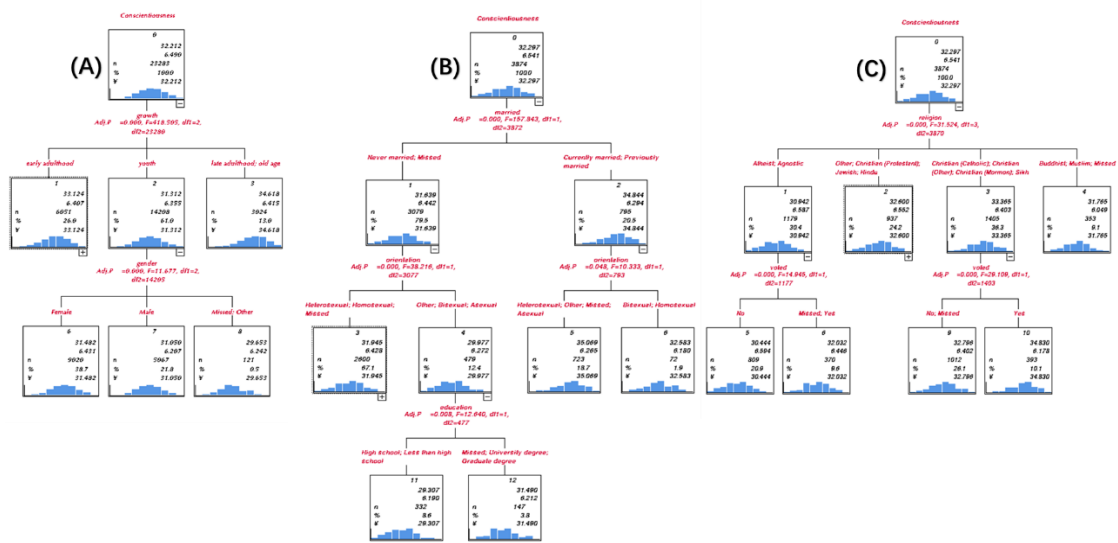

Fig. 6  Exhaustive CHAID decision tree models for C prediction predicted by
A: biological parameters; B: family factors; C: culture factors.

*Openness prediction model*

Females of both-handed dominants at their adulthood and old age (Node 9, $\overline{Y_O} = 38.40$) are more open while single handed dominants at their youth (Node 10, $\overline{Y_O} = 36.05$) are less open in personality (*Fig*.7-A). Bisexual orientation person with education of university degree who come from small size family (Node 15, $\overline{Y_O} = 38.88$) are more open in personality. While person with only high school education without specific sexual orientation (Node 6, $\overline{Y_O} = 34.75$) scored less in openness trait (*Fig*.7-B). Religious person of Atheists, Christians (Protestant), Jewish, Agnostic and Sikhs who voted (Node 5, $\overline{Y_O} = 37.46$) scored higher on the openness dimension. Muslims (Node 3, $\overline{Y_O} = 35.04$) score low on the openness dimension whether they vote or not (*Fig*.7-C).

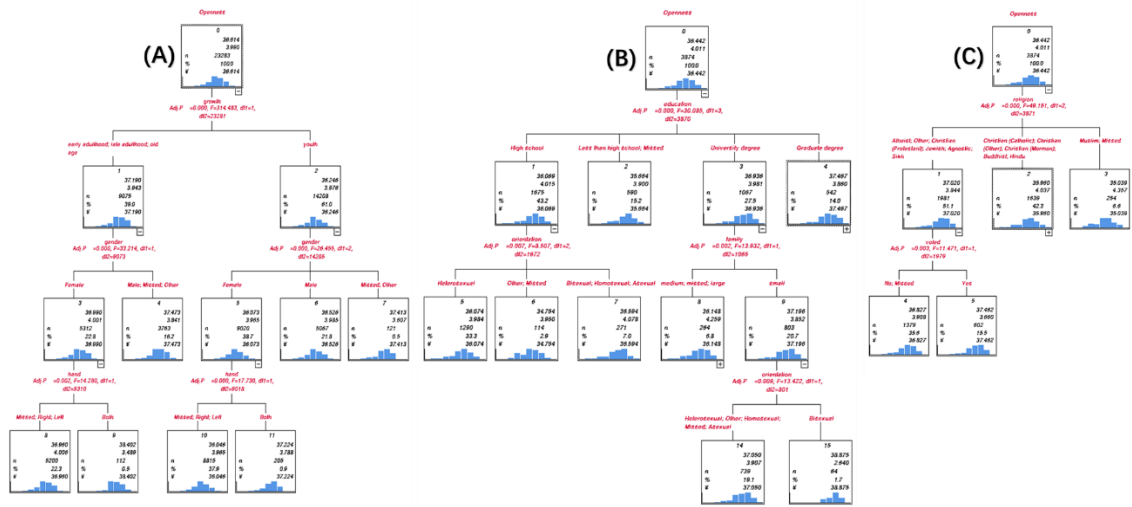

Fig. 7  Exhaustive CHAID decision tree models for O prediction predicted by
A: biological parameters; B: family factors; C: culture factors.

# Discussion

*Biological Factors*

**Gender differences related personality does exist based on statistic.** Females are more extroverted, neurotic and conscientious. And males are more agreeable and open. The evidence on gender differences in variance suggested that both genders gravitate towards their traditional roles just like the other research[8,9]. Cultural expectations about gender roles might form the stereotypes about gender differences[10]. Furthermore, in social role development theories, gender roles perception, gender socialization and sociostructurally power differentials are assumed to be the main reasons of gender differences. The fact that men and women are more similar than different in genetic, physical and psychological aspects is undeniably true. Scientists believed that trans-gender differences in personality exist partially from evolved psychological adaptations[11,12].

**Cross-age differences in personality traits are substantial.** The statistical evidence shows that persons at late adulthood and old age scored higher in Extroversion than early adulthood, and even much higher than youth. The youth scored higher in neuroticism than early adulthood, and those of early adulthood scored higher than late adulthood and old age. Same happened in the scoring of agreeableness dimension. People grow more extraverted but less neurotic when aging.

People at late adulthood and old age are more conscientious than at early adulthood, and much more than at youth. In the openness dimension, scored at adulthood and old age are higher than youth. They become less agreeable but more conscientious as grow older and become more open in personality since adulthood and stay open through the rest time of their lifespan. The continuity and change in the course of life consistent with the results of the Dutch participants of 7 years of assesment[13,14].

**Mechanisms of hand preference shaping personality still need to be further explored.**
Dominant hand does not determine a person's extraversion, neuroticism, or conscientiousness according to the statistical results of this study. But people who are accustomed to one-handed (left-handed or right-handed) are less agreeable and open in personality than those who use both hands. Previous studies have proved left and right-handers have differences in the cerebral functional organization and neural substrate[15]. Cooperative hand movements are controlled by neural coupling which involve the respective ipsilateral hemispheres[16]. But yet more evidences are needed to interpretate the link between personality trait and the and neural substrate.

*Family Factors*

**Sexual Orientation and variety in personality.** Heterosexuals are more extroverted than homosexual/bisexual, while homosexual/bisexual persons are more extroverted than asexual. Sexual orientation does not determine the neurotic aspect of personality. But in the agreeableness dimension, bisexual/asexual persons scored higher than heterosexual/homosexual ones. The bisexual/asexual people are not as good as those heterosexual/homosexual ones in the conscientious dimension. And the bisexual/homosexual persons are more open than heterosexual/asexual ones. Variety in personality of different sexual orientation person are assumed to relate to their experienced different social realities and developmental origins of sexual orientation[17,18]. Researches focusing on genes, anatomy and hormonal influences on psychosexual development still can't explain the controversy origins of sexual orientation[19]. As the important components of sexuality and sexual health, sexual orientation has proven theoretically predicted by personality[20]. This might help professionals developing services to promote sexually healthy.

**Native language.** According to the big 5 inventory survey, English native speakers are more extroverted and neurotic but are less agreeable than non-native English speakers. Besides, English native speakers are more conscientious and open. Personality is significantly related to native language used. This is very similar to previous research that concluded personality traits correlated with linguistic variables significantly[21]. Language interacts with personality constantly[22]. Personality differences will affect language learning[22]. Individuals' personality may significantly influence their verbal production[24]. In turn, individuals develop personality through natural language[25].

**Educational Attainment and Personality.** The statistics of this study show that people who have received university/graduate degrees are more extroverted than those received less than high school/high school degrees. The score on the neurotic dimension tends to decrease as the level of education received increases: less than high school>high school>university>graduate degree. But those who have received less than high school/high school degree are more agreeable in personality than those received university/graduate degree. In the conscientiousness and openness dimensions, the scores increase with the level of education received, and the data show that it is less than high school<high school<university<graduate degree. In summary, the data give us a personality profile of graduate degree, which is, extroverted, conscientious and open but not neurotic or always agreeable. Personality traits are closely related to education. The close correlation between personality traits and educational level can be explained to have relationships with the cognitive skills improvement and socioeconomic attainment according previous researchs[26,27].

**Marriage and Personality.** People who previously/currently married are more extroverted than those never married ones. Never married persons are more neurotic and agreeable in personality than those previously/currently married. Previously/currently married people scored more in conscientious and open dimensions than those never married ones. In conclusion, whether previously married or currently married, marriage seem to help shaping personality to be more extroverted, less neurotic or agreeable and more conscientious and open. That is mainly because marriage play positive role in promoting the consistency interindividual organization of personality attributes[28].

**Nurturing environment.** People raised in urban (town, city) are more agreeable than those who raised in suburban and rural (country side). People raised in urban and rural are more open in personality than those who raised in suburban. Personality in extroversion, neuroticism and conscientiousness dimensions are not distinguished clearly by raised environment when people grow. The results are consistent with the conclusion that certain characteristics of personality have impact on environmental choice[29].

**Family size.** People from small families are more neurotic than people from medium or large families. People from medium families are more conscientious than those from small or large families. The higher scores in openness dimension of a person comes from the smaller the size of the family. The order of the scores in the openness dimension is small>medium>large family. Family size has no distinguishing effect on extroversion and agreeableness dimensions of personality. Documentation reveals children from small families tend to accrue advantages in many developmental areas, especially intelligent[30]. Neuroticism and openness of dimensional personality might contribute to the growth advantage based on statistic results of this study.

*Culture Factor*
**Religion.** Some research revealed that cultural adaptation of agreeableness and conscientiousness can be partly explained individual differences in religiousness[31]. unlike these results, the influence between personality and religiousness in our study is not yet clear. But Jewish are more extroverted, neurotic and open, but less agreeable in personality. And the Atheist are less extroverted and conscientious, but more neurotic, agreeable and open in personality in this study. More to explore in the religiousness as a culture factor that could ultimately shape the personality.

**Political participant.** Voting is a participatory patterns individual engage in politics. Scientists have found that personality affiliates with particular type of political participant[32]. In this study, people who have participated in voting are more extroverted, conscientious and open but less neurotic and agreeable in personality than those who have not participated in voting. According to some social psychologist that personality traits potentially influence person's behavior by activating and stimulating people in interaction with the social environment.

**Interracial differences.** Races of Asian and Black scored higher than the Indigenous, Australian,

Native American / White and Arab in the personality dimension of Extroversion. Based on our statistic, races of Indigenous, Australian, Native American/White, Arab are more neurotic than Asian and Black. When it comes to Agreeableness in personality, Arab and Asian scored more than Indigenous, Australian, Native American/White and Black. The racial factor cannot distinguish the Conscientiousness dimension in this study. But in Openness dimension, Indigenous, Australian, Native American/White, Arab and Black scored higher than Asian. Although we know the characteristics of the various dimensions of racial personality, the pattern of personality differences between races still needs more evidence.

## Conclusion

This study analyzes and verifies through the analysis of data from three large-scale online big-5 personality questionnaires that the establishment of personality structure is not only affected by biological parameters internally, but also affect by external factors such as family and culture factors. Large-scale data is more conducive to help researchers see subtle and meaningful differences. Statistical analysis also supports the hypothesis that differences exist in general. Therefore, there is no doubt that personality traits are affected by multiple confounding factors. The exploration on multiplies cultures related religion, politics and race still needs more evidence. Nevertheless, this study has not found the which key factors that are crucial to the process of shaping of personality traits for specific populations. Future research directions should point to exploring more details of these meaningful influencing factors.